

\magnification\magstep1
\hoffset=0.5truecm
\voffset=0.5truecm
\hsize=15.8truecm
\vsize=23.truecm
\baselineskip=14pt plus0.1pt minus0.1pt \parindent=19pt
\lineskip=4pt\lineskiplimit=0.1pt      \parskip=0.1pt plus1pt


\def\Na{{\bf N}} 
\def\Qu{{\bf Q}} 
\def\Re{{\bf R}} 
\def\Ze{{\bf Z}} 
\def\A{{\cal A}}
\def\Anul{{\cal A}_0}
\def\ua{{\rm C}^*({\cal A})}

\def\B{{\cal B}}

\def\D{\Delta}

\def\H{{\cal H}}

\def\k{\kappa}
\def\K{{\cal K}}

\def\l{\lambda}
\def\L{\Lambda}

\def\o{\omega}
\def\O{{\cal O}}

\def\Q{\Omega}
\def\r{\rho}

\def\s{\sigma}

\def\S{{\cal S}}
\def\t{\tau}

\def\W{{\cal W}}

\def\ad{{\rm ad}}
\def\Pp{{\cal P}_+}                     
\def\Ppo{{\cal P}_+^\uparrow}           

\def\imply{\Rightarrow}
\def\np{\par\noindent}

\font\sectionfont=cmbx10 scaled\magstep1
\def\titlea#1{\vskip0pt plus.3\vsize\penalty-75
    \vskip0pt plus -.3\vsize\bigskip\bigskip
    \noindent{\sectionfont #1}\nobreak\smallskip\noindent}
\def\titleb#1{\medskip\noindent{\it#1}\qquad}
\def\claim#1#2{\vskip.1in\medbreak\noindent{\bf #1.} {\sl #2}\par
    \ifdim\lastskip<\medskipamount\removelastskip\penalty55\medskip\fi}
\def\rmclaim#1#2{\vskip.1in\medbreak\noindent{\bf #1.} {#2}\par
    \ifdim\lastskip<\medskipamount\removelastskip\penalty55\medskip\fi}
\def\beglemma#1#2\endlemma{\claim{#1 Lemma}{#2}}
\def\begdefinition#1#2\enddefinition{\claim{#1 Definition}{#2}}
\def\begtheorem#1#2\endtheorem{\claim{#1 Theorem}{#2}}
\def\begcorollary#1#2\endcorollary{\claim{#1 Corollary}{#2}}
\def\begremark#1#2\endremark{\rmclaim{#1 Remark}{#2}}
\def\begproposition#1#2\endproposition{\claim{#1 Proposition}{#2}}
\def\begassumption#1#2\endassumption{\claim{#1}{#2}}
\def\begProof{\noindent{\bf Proof.}\quad}
\def\begProofof#1{\medskip\noindent{\bf Proof of #1.}\quad}
\def\square{\hbox{$\sqcap\!\!\!\!\sqcup$}}
\def\endProof{\hfill\square\par
    \ifdim\lastskip<\medskipamount\removelastskip\penalty55\medskip\fi}
\newcount\FNOTcount \FNOTcount=1

\def\addfnot{\global\advance\FNOTcount by 1}

\def\refno#1#2{\item{[#1]}{#2}}
\def\begref#1#2{\titlea{#1}}
\def\endref{}
\def\np{\par\noindent}
\def\TitBib{References}

\newcount\REFcount \REFcount=1
\def\numref{\number\REFcount}
\def\addref{\global\advance\REFcount by 1}
\def\wdef#1#2{\expandafter\xdef\csname#1\endcsname{#2}}
\def\wdch#1#2#3{\ifundef{#1#2}\wdef{#1#2}{#3}
    \else\write16{!!doubly defined#1,#2}\fi}
\def\wval#1{\csname#1\endcsname}
\def\ifundef#1{\expandafter\ifx\csname#1\endcsname\relax}

\def\autonumref{
    \def\rfr(##1){\wdef{q##1}{yes}\ifundef{r##1}$\diamondsuit$##1
        \write16{!!ref ##1 was never defined!!}\else\wval{r##1}\fi}
    \def \REF(##1)##2\endREF{\wdch{r}{##1}{\numref}\addref}\REFERENCES
    \def\references{
        \def \REF(####1)####2\endREF{
            \ifundef{q####1}\write16{!!ref. [####1] was never quoted!!}\fi
            \refno{\rfr(####1)}####2}
        \begref{\TitBib}{99}\REFERENCES\endref}}

\def\REFERENCES{
 \REF(BiWi1) Bisognano J., Wichmann E., ``{\it On the duality condition for a
Hermitian scalar field}'', J. Math. Phys. {\bf 16} (1975), 985-1007.
 \endREF
 \REF(BiWi2) Bisognano J., Wichmann E., ``{\it On the duality condition for
quantum fields}'', J. Math. Phys. {\bf 17} (1976), 303-321.
 \endREF
 \REF(Borc1) Borchers H.J., ``{\it The CPT theorem in two-dimensional theories
of local observables}'', Commun. Math. Phys. {\bf 143} (1992), 315.
 \endREF
 \REF(BorcP) Borchers H.J., ``{\it On the use of modular groups in quantum
field
theory}'' Ann. Inst. H. Poincar\'e, this Issue.
 \endREF
 \REF(BGLo1) Brunetti R., Guido D., Longo R., ``{\it Modular structure and
duality in conformal quantum field theory}'', Commun. Math. Phys., {\bf 156}
(1993), 201-219.
 \endREF
 \REF(BGLo2) Brunetti R., Guido D., Longo R., ``{\it Group cohomology, modular
theory and space-time symmetries}'', Rev. Math. Phys., {\bf 7} (1994),
57-71.
 \endREF
 \REF(BGLo3) Brunetti R., Guido D., Longo R., in preparation.
 \endREF
 \REF(BuSu1) Buchholz D., Summers S.J., ``{\it An algebraic characterization of
vacuum states in Minkowski space}'', Commun. Math. Phys. {\bf 155} (1993),
442-458.
 \endREF
 \REF(Davd1) Davidson D., ``{\it Endomorphism Semigroups and Lightlike
Translations}'', preprint (hep-th/9412094).
 \endREF
 \REF(Davd2) Davidson D., in preparation
 \endREF
 \REF(DHRo1) Doplicher S., Haag R., Roberts J.E., ``{\it Local observables and
particle statistics I}'', Commun. Math. Phys. {\bf 23} (1971), 199-230.
 \endREF
 \REF(DHRo2) Doplicher S., Haag R., Roberts J.E., ``{\it Local observables and
particle statistics II}'', Commun. Math. Phys. {\bf 35} (1974), 49-85.
 \endREF
 \REF(DoRo1) Doplicher S., Roberts J.E., ``{\it Why there is a field algebra
with a compact gauge group describing the superselection structure in particle
physics}'', Commun. Math. Phys. {\bf 131} (1990), 51-107.
 \endREF
 \REF(Fred1) Fredenhagen K., ``{\it Generalization of the theory of
superselection sectors}'', in The algebraic Theory of Superselection Sectors,
D.
Kastler ed., World Scientific, Singapore 1990.
 \endREF
 \REF(GaFr1) Gabbiani F., Fr\"ohlich J., ``{\it Operator algebras and Conformal
Field Theory}'', Commun. Math. Phys. {\bf 155} (1993), 569-640.
 \endREF
 \REF(GuLo1) Guido D., Longo R., ``{\it Relativistic invariance and charge
conjugation in quantum field theory}'', Commun. Math. Phys. {\bf 148} (1992),
521-551.
 \endREF
 \REF(GuLo2) Guido D., Longo R., ``{\it An algebraic Spin and Statistics
theorem. I}'', to appear in Commun. Math. Phys.
 \endREF
 \REF(GuLo3) Guido D., Longo R., ``{\it The Conformal Spin and Statistics
theorem}'', preprint (hep-th/9505059).
 \endREF
 \REF(HiLo1) Hislop P., Longo R., ``{\it Modular structure of the local
algebras
associated with the free massless scalar field theory}'', Commun. Math. Phys.
{\bf 84} (1982), 84.
 \endREF
 \REF(Jors2) J\"orss M., ``{\it The construction of pointlike fields from
conformal Haag-Kastler nets}'', preprint (hep-th/9506016)
 \endREF
 \REF(Kuck2) Kuckert B., ``{\it A new approach to Spin \& Statistics}'',
preprint (hep-th/9412130).
 \endREF
 \REF(Kuck3) Kuckert B., ``{\it Borchers' commutation relations and modular
symmetries}'', \np\qquad preprint.
 \endREF
 \REF(Long2) Longo R., ``{\it Index of subfactors and statistics of quantum
fields. I}'', Commun. Math. Phys. {\bf 126} (1989), 217-247.
 \endREF
 \REF(Long3) Longo R., ``{\it Index of subfactors and statistics of quantum
fields. II}'', Commun. Math. Phys. {\bf 130} (1990), 285-309.
 \endREF
 \REF(OkTo1) Oksak A.I., Todorov I.T., ``{\it Invalidity of TCP-theorem for
infinite-component fields}'', Commun. Math. Phys. {\bf 11} (1968), 125-130.
 \endREF
 \REF(Wies2) Wiesbrock H.V., ``{\it Conformal quantum field theory and
half-sided modular inclusions of von~Neumann algebras}'', Commun. Math. Phys.
{\bf 158}, (1993) 537-544.
 \endREF
 \REF(Yngv1) Yngvason J., ``{\it A note on essential duality}", Lett. Math.
Phys. {\bf31} (1994) 127-141.
 \endREF}

\autonumref


\topskip3.cm
\font\ftitle=cmbx12 scaled\magstep1
\centerline{\ftitle Modular Covariance, PCT, Spin and Statistics}

 \bigskip\bigskip\bigskip

\centerline{Daniele Guido\footnote{$^*$}
{Supported in part by MURST and CNR-GNAFA.}}
\footnote{}{E-mail:\ guido@mat.utovrm.it}
 \bigskip\bigskip\bigskip

Dipartimento di Matematica, Universit\`a di Roma ``Tor Vergata''

via della Ricerca Scientifica, I--00133 Roma, Italia

\vskip2.cm
\hfill June 1995\vskip2.cm
\noindent
 {\bf Abstract.} The notion of modular covariance is reviewed and the
reconstruction of the Poincar\'e group extended to the low-dimensional case.

The relations with the PCT symmerty and the Spin and Statistics theorem are
described.

 \vfill\eject


\topskip0.cm

\titlea{Introduction}

The first relation between some space-time transformations and the
modular group of the von~Neumann algebras associated with wedge
regions was discovered by Bisognano and Wichmann in the particular
case of Wightman fields [\rfr(BiWi1),\rfr(BiWi2)]. They also proved
that the modular conjugation implements both the space-time
reflection w.r.t. the edge of the wedge and the charge conjugation.
The analogous result for conformally covariant theories was then
proven in [\rfr(HiLo1)].

Since the Bi\-so\-gna\-no-Wich\-mann relations are far more intrinsic in the
algebraic setting than in the Wightman field approach, it is very natural to
conjecture such relations to hold  for the local algebras of a quantum field
theory under general hypotheses.

Indeed the examples in [\rfr(OkTo1)] and in [\rfr(Yngv1)], where the modular
operators are not associated with some covariant representation of the
space-time symmetries of the theory, suggest that some kind of compactness
condition (e.g. the split property) and Poincar\'e covariance could give such
general hypotheses, but this conjecture is far from being proven.

In 1992, Borchers proposed a different approach to the problem.
Instead of showing the Bi\-so\-gna\-no-Wich\-mann relations for a
given covariant representation, his purpose was to reconstruct the
space-time symmetries in terms of the modular operators. In
particular he showed that for positive energy, translationally
covariant, one- or two-dimensional theories, modular groups and
translations give a covariant representation of the Poincar\'e
group [\rfr(Borc1)].

The theorem of Borchers had several consequences. In particular we
quote the solution of the Bi\-so\-gna\-no-Wich\-mann conjecture for
conformal theories [\rfr(BGLo1),\rfr(GaFr1)], and the
characterization of the conformal theories on $S^1$ in terms of
half-modular inclusions [\rfr(Wies2)]. Converses of Borchers
theorem are contained in [\rfr(Davd1)].

Borchers' purpose may be pursued in terms of a different set of
hypotheses, namely the geometrical meaning of some modular objects.

 Buchholz and Summers [\rfr(BuSu1)] were able to reconstruct the
translation group assuming that the modular conjugations of wedge
regions implement space-time reflections [see also [\rfr(BorcP)].

 The reconstruction of the whole group of space-time symmetries
for high-di\-men\-sio\-nal theories is studied in [\rfr(BGLo2)].
Neither translation covariance nor essential duality is assumed
there, but the essence of the Bi\-so\-gna\-no-Wich\-mann
prescription: the modular groups of the wedge regions should
implement the correct space-time transformations. Such an
assumption, which was called modular covariance, is sufficient to
reconstruct a covariant representation of the Poincar\'e group and
imply the second Bi\-so\-gna\-no-Wich\-mann property, namely the
relation between modular conjugations and space-time reflections.

A related result in [\rfr(Kuck3)] shows that if the modular
conjugation, resp. the modular group (dimension $\geq4$),
implement a geometrical transformation whatsoever, then it
implements the correct one.
 A classification of different, generalized forms
of modular covariance is contained in [\rfr(Davd2)].

Up to now conformal theories were the only place where both
approaches completely solve the problem. Indeed Borchers theorem
implies, as already mentioned, that conformal theories verify the
Bi\-so\-gna\-no-Wich\-mann properties, while (conformal) modular covariance
reconstructs the unique covariant, positive-energy representation of
the conformal group ([\rfr(BGLo2)], see also Remark~1.6). On the
other hand, when the reconstruction of the Poincar\'e group is
concerned, modular covariance was confined to the high dimensional
case and Borchers technique to the low dimensional one.

Here we show that the modular covariance assumption may
reconstruct the Poin\-ca\-r\'e group in the one and two dimensional
case either (cf. Section~1). Moreover, for any space-time
dimension, such assumption may be weakened to a more intrinsic one
if essential duality is assumed. More precisely, only the covariant
action of the modular automorphism group associated with a wedge
$W$ on the algebras of subregions of $W$ is requested.

In this way positive energy translational covariance is equivalent
to (weak) modular covariance for low-di\-men\-sio\-nal theories
satisfying essential duality.

As we mentioned before, the results of Bisognano and
Wichmann are intimately tied with the PCT symmetry and therefore
with the Spin and Statistics relation. Conversely, modular
covariance properties give sufficient hypotheses for the PCT and
Spin and Statistics theorems to hold, as it is shown in
[\rfr(GuLo1)], [\rfr(GuLo2)], [\rfr(Kuck2)], [\rfr(GuLo3)],
[\rfr(BGLo3)]. We present and discuss some of these results in
Section~2.

\titlea{1. Modular covariance on the Minkowski space}

In this section we review some results about modular covariance
studied in [\rfr(BGLo2)] and extend such results to
low-di\-men\-sio\-nal Minkowski spaces, thus making them comparable
with the theorem of Borchers about the Bi\-so\-gna\-no-Wich\-mann
property and PCT  symmetry in the two-dimensional space-time
[\rfr(Borc1)], see also [\rfr(BorcP)].

Moreover we weaken the modular covariance assumption, requesting
only that the intrinsic action of the modular automorphism group of
a wedge on the algebras associated with some of its subregions has
the prescribed geometrical meaning. As a counterpart, we assume
essential duality, which follows when the stronger modular
covariance is assumed [\rfr(BGLo1)].

We shall always consider local precosheaves on the wedges of the
$n$-dimensional Minkowski space $M$, $n\geq1$, i.e. maps
 $$
W\to\A(W)
 $$
from the family $\W$ of wedge regions in $M$ (when $n=1$ wedge
regions are, by definition, open half lines) to von~Neumann algebras
on a separable Hilbert space $\H$ verifying the {\it isotony}
property:
 $$
 W_1\subset W_2\imply \A(W_1)\subset\A(W_2)
 $$
 and the stronger form of locality called {\it essential duality}:
 $$
 \A(W')=\A(W)'\ ,
 $$
 where $W'$ denotes the space-like complement of $W$ (the interior
of the complement when $n=1$).

We denote by $\Pp$, resp. $\Ppo$, the proper, resp. proper
orthochronous Poincar\'e group. If $n=1$, $\Pp$, resp. $\Ppo$ is
the group of affine, resp. orientation preserving affine
transformations.

If $W\in\W$, $\L_W$ denotes the one-parameter group of boosts
preserving $W$ (no rescaling is adopted here). When $n=1$ and $W$ is a right
half-line $\L_W$ is
the one parameter group of dilations fixing the edge of $W$.
$\L_W$ for left half lines is determined by the equation
$\L_W(t)=\L_{W'}(-t)$.
 We also denote by $r_W$ the element in $\Pp$ corresponding to the
reflection w.r.t. the edge of $W$.

The main result of this section is to show that under the weak
modular covariance assumption (see below) we may construct a
canonical representation of the proper Poincar\'e group acting on
the local algebras consistently with the action of $\Pp$ on $M$.
In particular this gives a PCT operator, i.e. an antiunitary
operator which corresponds to the PT transformation on $M$ and
implements the charge conjugation on superselection sectors (cf.
[\rfr(GuLo1),\rfr(GuLo3)]).

First we discuss  the low-di\-men\-sio\-nal case ($n=1,2$), which is
interesting in itself and furnishes the basis for the weakening
of the modular covariance assumption in the higher dimensional
Minkowski space.

\titleb{The low-dimensional case, $n\leq2$.}

\begtheorem{1.1} Let $\A$ be a precosheaf on the wedges of the
$n$-dimensional Minkowski space, $n\leq2$, satisfying essential
duality.
 Assume also the existence of a vector $\Q$ ({\it vacuum})
cyclic for the algebras of all wedges, and {\it weak modular
covariance}: if $W_1\supset W_2$
then
 $$
 \s_{W_1}^t(\A(W_2))=\A(\L_{W_1}(2\pi t)W_2)\eqno(1.1)
 $$
where $\s_W$ denotes the modular automorphism group of the algebra
$\A(W)$  associated with the state $\o:=(\Q,\cdot\ \Q)$.
 \np
 Then there is a positive energy (anti)-unitary
representation $U$ of the Poincar\'e group $\Pp$ determined by
 $$
\eqalign{
U(\L_W(t))&=\D_W^{{it\over2\pi}}\cr
U(r_W)&=J_W}
 $$
The representation $U$ implements precosheaf maps, i.e.
 $$
U(g)\A(W)U(g)^*=\A(gW)\ ,\qquad g\in\Pp\ .\eqno(1.2)
 $$
 \endtheorem

First we prove some lemmas concerning the the one-dimensional case,
and adopt the subscript $a$ when dealing with objects associated
with the right half line (wedge) $(a,+\infty)$.
 First we observe that when $a\leq b$, weak modular covariance for
the inclusion $\A(a,+\infty)$ $\supset$ $\A(b,+\infty)$ implies
$\D_a^{it}\D_b^{is}\D_a^{-it}=\D_{\L_a(2\pi t)b}^{is}$ . On the
other hand, essential duality implies
 $$
 \D_{(-\infty,a)}^{it}=\D_a^{-it}\ ,\qquad J_{(-\infty,a)}=J_a\ .
 \eqno(1.3)
 $$
 If $a>b$, we apply weak modular covariance for the inclusion
$\A(-\infty,a)$ $\supset$ $\A(-\infty,b)$. Therefore we immediately
get the following.
 \beglemma{1.2} For all $a,b,s,t\in\Re$ we have
 $$
 \D_a^{it}\D_b^{is}\D_a^{-it}=\D_{\L_a(2\pi t)b}^{is}\ .
 $$
 \endlemma

Next step will be the construction of the translation group (cf.
[\rfr(BorcP)], Theorem~7.2 for an analogous construction). To this
aim, let us consider the map from $\Re$ to the unitaries on $\H$
given by
 $$a\to T(a):=\D_0^{is_0}\D_a^{-is_0}\ ,$$
 where $s_0:={\log2\over2\pi}$.

\begproposition{1.3} The map $a\to T(a)$ is a strongly continuous
one-parameter group that implements translations, namely
 $$
T(a)\A(W)T(a)^*=\A(W+a)\ .\eqno(1.4)
 $$
 \endproposition

\begProof A direct application of Lemma~1.2 and of the
definition of $T(a)$ gives
 $$
\eqalign{
 T(a)\D_b^{it}T(a)^*&=\D_{b+a}^{it}   \cr
 T(a)^*\D_b^{it}T(a)&=\D_{b-a}^{it}\ .\cr}\eqno(1.5)
 $$
Now we show that
 $$
 T(a)^n=T(na)\ ,\qquad n\in\Ze\eqno(1.6)
 $$
 First we observe that, by (1.5), for any $n\in\Na$,
 $$
T(a)=T(a)^n\D_0^{is_0}\D_a^{-is_0}T(a)^{-n}
=\D_{na}^{is_0}\D_{(n+1)a}^{-is_0}\ . \eqno(1.7)
 $$
 Then, assuming (1.6) for a given $n$ and making use of (1.7)
we have
 $$
 T(a)^{n+1}= T(na)T(a)=\D_0^{is_0}\D_{na}^{-is_0}
\D_{na}^{is_0}\D_{(n+1)a}^{-is_0}=T((n+1)a)\ ,
 $$
 and (1.6) for $n\in\Na$ follows by induction. Since, by (1.5),
 $$
\eqalign{T(-a)
&=T(-a)^*\D_0^{is_0}\D_{-a}^{-is_0}T(-a)\cr
&=T(-a)^*\D_0^{is_0}T(-a)\ T(-a)^*\D_{-a}^{-is_0}T(-a)
=T(a)^*\ ,\cr} \eqno(1.8)
 $$
equation (1.6) holds for negative $n$ too.
 \np
 Equation (1.6) immediately imply that
 $$
\forall a,b\in\Qu,\qquad T(a)T(b)=T(a+b)\ .\eqno(1.9)
 $$
 Finally by Lemma~1.2 we have, if $|a|<1$,
 $$
 \D_0^{is_0}\left(
\D_0^{{i\over2\pi}\log(a+1)}\D_1^{-{i\over2\pi}\log(a+1)}
\right)\D_0^{is_0}\left(
\D_0^{{i\over2\pi}\log(a+1)}\D_1^{-{i\over2\pi}\log(a+1)}
\right)^*=T(a)
 $$
 which shows that $T(a)$ is strongly continuous in a neighborhood
of the origin therefore, by relation~(1.9), it is a
strongly continuous one-parameter group.
 \np
 Now we prove (1.4). If $b\geq0$ we have
 $$
T(b)\A(a,+\infty)T(b)^*
=\D_{a-b}^{is_0}\D_a^{-is_0}\A(a,+\infty)\D_a^{is_0}\D_{a-b}^{-is_0}
=\A(a+b,+\infty)
 $$
 where we have used $T(b)$ $=$
$T(a-b)\D_0^{is_0}\D_b^{-is_0}T(a-b)^*$ $=$
$\D_{a-b}^{is_0}\D_a^{-is_0}$ and applied weak modular covariance
for the inclusion $(a-b,+\infty)$ $\supseteq$ $(a,+\infty)$. If
$b<0$ we may write $T(b)$ as $\D_{a+b}^{is_0}\D_{a+2b}^{-is_0}$ and
then apply weak modular covariance for the inclusion
$(a+2b,+\infty)$ $\supseteq$ $(a,+\infty)$. Essential duality
implies (1.4) for left wedges too.\endProof

\begProofof{Theorem~1.1} THE ONE-DIMENSIONAL CASE. First we
show that the modular unitary groups associated with the wedge
algebras generate a representation of $\Ppo$. Since we already
proved relation~(1.5) and $\Ppo$ is the semidirect product of
translations and dilations, it is sufficient to show that, for all
$a,b\in\Re$  we have
 $$
 \D_0^{it}T(a)\D_0^{-it}=T(e^{2\pi t}a)\eqno(1.10)
 $$
 Set $c_t(a)$ $=$
$\D_0^{it}T(a)\D_0^{-it}T(e^{2\pi t}a)^*$.
By definition of $T$ and by a repeated application of Lemma~1.2 we
check that $c_t(a)$ commutes with $\D_b^{is}$, $\forall
a,b,s,t\in\Re$. As a consequence
 $$
\eqalign{c_t(a)c_t(b)
&=\D_0^{it}T(a)\D_0^{-it}c_t(b)T(e^{2\pi t}a)^*\cr
&=\D_0^{it}T(a+b)\D_0^{-it}T(e^{2\pi t}(a+b))^*
=c_t(a+b)\cr}
 $$
 and, exploiting the dependence on $t$,
 $$
 \eqalign{c_{t+s}(a)
&=\D_0^{it}
\left(c_s(a)T(e^{2\pi s}a)\right)\D_0^{-it}T(e^{2\pi(t+s)}a)^*\cr
&=c_s(a)c_t(e^{2\pi s}a)
=c_t((e^{2\pi s}-1)a)c_t(a)c_s(a)\ ,\cr}
 $$
where we used the centrality of $c_t(a)$ in the group generated by the modular
unitaries and the multiplicativity of $c_t(\cdot)$.
 Then, interchanging $t$ with $s$ in the previous equation, we get
$c_t((e^{2\pi s}-1)a)=c_s((e^{2\pi t}-1)a)$ hence, recalling that
$s_0={\log2\over2\pi}$,
 $$c_t(a)=c_{s_0}((e^{2\pi t}-1)a)\eqno(1.11)$$
 Now, making use of equations~(1.5) and (1.8) we obtain
 $$
\eqalign{c_{s_0}(a)
&=\D_0^{is_0}T(a)\D_0^{-is_0}\ \D_0^{is_0}\D_{-2a}^{-is_0}\cr
&=\D_0^{is_0}\D_{-a}^{-is_0}T(a)=1\cr}\qquad\forall t\in\Re
 $$
 hence, by (1.11), $c_t(\cdot)\equiv1$, namely (1.10) holds.
 \np
 An analytic continuation argument (see e.g. [\rfr(BGLo2)],
Proposition~2.7) shows that the generator of the translations is
positive, therefore the theorem of Borchers ([\rfr(Borc1)], see also
[\rfr(BorcP)]) applies and we get $J_0T(a)J_0=T(-a)$. This relation
immediately imply that, setting $U(r_0)=J_0$, we get an
(anti)-unitary representation of $\Pp$.
 \np
We conclude this proof showing that $U(g)$ implements a precosheaf
map for any $g\in\Pp$. Since we already proved (1.4), it is
sufficient to show (1.2) only for $g=\L_0^{it}$, $t\in\Re$ and
for $g=r_0$. We prove the relation when $g=\L_0(t)$ and
$W=(a,+\infty)$ is a right wedge. The proofs for the other cases
are analogous.
 $$
\eqalign{
U(\L_0(t))\A(a,+\infty)U(\L_0(t))^*
&=\D_0^{{it\over2\pi}}T(a)\A(0,+\infty)T(a)^*\D_0^{-{it\over2\pi}}\cr
&=T(\L_0(t)a)\A(0,+\infty)T(\L_0(t)a)^*=\A(\L_0(t)a,+\infty)\cr}
 $$
\np THE TWO-DIMENSIONAL CASE.
 First we construct a positive energy, covariant representation of
the group of translations. Set $\t^\pm$ $=$ $\t(a,\pm a)$ where
$\t(v)$, $v\in\Re^2$, denotes the translation by $v$ on $\Re^2$ and
the first coordinate is the time coordinate, and set
$W_a^+:=\t^+(a)W_0$, $W_a^-:=\t^-(a)W_0'$, where $W_0$ is the right
wedge whose edge is the origin. Then we may consider the
one-dimensional sub-precosheaves given by $(\A(W_a^+),\A(W_a^+)')$
and by $(\A(W_a^-),\A(W_a^-)')$. For such one-dimensional
precosheaves weak modular covariance holds, hence, by
Proposition~1.3, we get two one-parameter translation groups
$T^\pm(a)$ such that $T^\pm(a)\A(W_b^\pm)T^\pm(a)^*$ $=$
$\A(W_{a+b}^\pm)$. Recalling equations (1.3) and (1.10), we get
 $$
\D_{W_0}^{it}T^\pm(a)\D_{W_0}^{-it}=T^\pm(e^{\pm2\pi t}a)\ .\eqno(1.12)
 $$
 First we show that these two light-like translations implement
precosheaf maps, i.e.
 $$
\eqalignno{
T^+(a)\A(W)T^+(a)^*&=\A(\t^+(a)W)&(1.13)\cr
T^-(a)\A(W)T^-(a)^*&=\A(\t^-(a)W)&(1.14)\cr}
 $$
Observe that if $W_1\supseteq W_2\supseteq W_3$ then, for each
$t\in\Re$,
 $$
\D_1^{it}\D_2^{-it}\A(W_3)\D_2^{it}\D_1^{-it}
=\A(\t((\L_1(2\pi t)-1)v)\ W_3)\eqno(1.15)
 $$
where $\D_i^{is}$, resp. $\L_i(s)$ denotes the modular unitary
group, resp. the one parameter group of boosts associated with the
wedge $W_i$, $i=1,2$ and $v\in\Re^2$ is defined by $W_2=W_1+v$.
 Let $(t,x)\in\Re^2$ be the edge of $W$. If $W$ is a right wedge and
$x\geq t$, we may apply equation (1.15) to the inclusion
$W_{t-2|a|}^+$ $\supseteq$  $W_{t}^+$ $\supseteq$ $W$, $a\in\Re$.
Since
 $$
T^+(a)=\D_{W^+_{t-2|a|}}^{is(a)}\D_{W^+_{t}}^{-is(a)}
 $$
where $s(a)=(2\pi)^{-1}\log3/2$ if $a\geq0$ and
$s(a)=(2\pi)^{-1}\log1/2$ if $a<0$, we get (1.13).
 If $W$ is a left
wedge and $x\geq t$, then $W_{t-2|a|}^+$ $\supset$  $W_{t}^+$
$\supset$ $W'$, therefore (1.13) for $W$ follows dualizing the
analogous relation for $W'$. Finally if $x<t$ we may consider
the inclusion $(W_{t+2|a|}^+)'$ $\supset$  $(W_{t}^+)'$ $\supset$
$W$ when $W$ is a left wedge or the inclusion $(W_{t+2|a|}^+)'$
$\supset$  $(W_{t}^+)'$ $\supset$ $W'$ when $W$ is a right wedge,
and (1.13) again follows. Equation (1.14) is proven in the same way.
 \np
Relations (1.13) and (1.14) implies that the multiplicative
commutator
 $$
c(s,t):=T^+(s)T^-(t)T^+(-s)T^-(-t)\eqno(1.16)
 $$
commutes with $\D_W^{it}$ for any $W\in\W$, $t\in\Re$. Now we show
that $c(t,s)\equiv1$. On the one hand, recalling equation~(1.12),
we get
 $$
c(s,t)=\D_{W_0}^{ir}c(s,t)\D_{W_0}^{ir}
=c(e^{2\pi r}s,e^{-2\pi r}t)\ .\eqno(1.17)
 $$
 On the other hand, multiplying equation (1.16) by
$(T^+(s)T^-(t))^*$ on the left and by $T^+(s)T^-(t)$ on the right
we get $c(s,t)=c(-s,-t)$. This, together with (1.17), implies that
$c(s,t)$ depends on only one variable:
 $$c(s,t)=c(1,st)=:c(st)\ .\eqno(1.18)$$
 Now, a direct computation show that $c(t)$ is a one parameter
group. Then equation (1.16) reads
 $$T^+(s)T^-(t)=c(st)T^-(t)T^+(s)\ ,$$
 namely we get a representation of the Heisenberg group. Since
the generators of $T^+(\cdot)$ and $T^-(\cdot)$ are positive,
$c(\cdot)\equiv1$. Indeed decomposing the representation of the
Heisenberg group along its center, any irreducible direct summand
for which $c(t)\not=1$ would give a representation of
the canonical commutation relations with positive generators,
which is impossible.
 \np
The rest of the proof is completely analogous to the
one-dimensional case.\endProof

\begremark{1.4} First we compare our result with that of Borchers
[\rfr(Borc1)] and observe that, for a precosheaf on the wedges of
the 1 or 2 dimensional space-time satisfying essential duality,
weak modular covariance is equivalent to the existence of positive
energy translations, and both assumptions imply the thesis of
Theorem~1.1.
 \np
Indeed we used Borchers theorem in the proof of Theorem~1.1 to
show that the relation $J_0T(a)J_0=T(-a)$ holds in the
one-dimensional case. We can give an alternate proof of this
relation. Analogously to [\rfr(GuLo2), Proposition~2.6], the
operator $\D_0^{1/2}T(a)\D_0^{-1/2}$ is densely defined and
extended by $J_0T(a)J_0$, therefore it is sufficient to check the
relation
 $$\D_0^{1/2}T(a)\D_0^{-1/2}\subset T(-a)\eqno(1.19)$$
in any positive energy representation of $\Ppo$.
 On the other hand, if $H$ is the generator of $T(a)$, the
one-parameter groups $t\to\D_0^{{it\over2\pi}}$ and $s\to e^{is\log
H}$ give a representation of the Weyl commutation relations.
Therefore since all representations of these commutation relations
are multiples of the Schr\"odinger representation, it is sufficient
to check relation (1.19) in one representation, e.g. the free field
of mass $m$, where it is known to hold.
 \np
 We now define the algebras associated with double cones (open
intervals in the one-dimensional case):
 $$\A(\O):=\bigcap_{W\supset\O}\A(W)\ .$$
The kernel of the representation $U$ is either a group of light-like
translations (in a two-dimensional theory) or the whole group of translations.
In the first case we get a one-dimensional theory as a degenerate case of a
two-dimensional one, in the latter the precosheaf consists of only two
algebras, $\A(W_R)$ and $\A(W'_R)=\A(W_R)'$, where $W_R$ is (any) right wedge,
and the algebras of all double cones coincide with the center of $\A(W_R)$. In
particular local algebras may be trivial.
 Conversely, if the algebra $\vee_t T^\pm(t)\A(\O)T^\pm(t)^*$ is
irreducible, then $\Q$ is cyclic for $\A(\O)$ (Reeh-Schlieder
theorem) and this implies that the algebras of wedge regions are
generated by the algebras of double cones. In this case, either $U$
is injective, or $\Ppo$ is in the kernel of $U$ and all the algebras $\A(W)$
coincide with a Maximal Abelian Sub-Algebra of $\B(\H)$ (cf. [\rfr(BGLo2)]).
\endremark

\titleb{The $n$-dimensional case, $n\geq3$.} Here we consider
a precosheaf on wedges of von~Neumann algebras in $\B(\H)$, $\H$ a
separable Hilbert space, and assume essential duality. Then set
 $$\A(\O):=\bigcap_{W\supset\O}\A(W)\qquad\O\in\K$$
 where $\K$ denotes the family of double cones in $M$, and require
that there exists a common cyclic vector $\Q$ for the algebras
associated with double cones. The weak modular covariance
assumption now takes this form:
 $$
 \s_{W}^t(\A(\O))=\A(\L_{W}(2\pi t)\O)\eqno(1.20)
 $$
where $\s_W$ denotes the modular automorphism group of the algebra
$\A(W)$  associated with the state $\o:=(\Q,\cdot\ \Q)$.

\begcorollary{1.5} Let $\A$ be a precosheaf on the wedges of the
$n$-dimensional Minkowski space satisfying the
mentioned properties.
 \np
 Then there is a positive energy (anti)-unitary
representation $U$ of the Poincar\'e group $\Pp$ determined by
 $$
\eqalign{
U(\L_W(t))&=\D_W^{{it\over2\pi}}\cr
U(r_W)&=J_W}
 $$
The representation $U$ implements precosheaf maps, i.e.
 $$
U(g)\A(W)U(g)^*=\A(gW)\ ,\qquad g\in\Pp\eqno(1.21)
 $$
 and, as a consequence,
 $$
U(g)\A(\O)U(g)^*=\A(g\O)\ ,\qquad g\in\Pp\ .
 $$
\endcorollary

\begProof First we notice that, for each $\O\in\K$, $\O\subset W$,
 $$
\bigvee_{t\in\Re}\A(\L_W(t)\O)=\A(W)
 $$
because the first von~Neumann algebra is a globally invariant
subalgebra of the latter w.r.t. the modular group, and admits $\Q$
as a cyclic vector. Therefore, whenever $W_1\supset W_2$, weak
modular covariance implies
 $$
 \s_{W_1}^t(\A(W_2))=\A(\L_{W_1}(2\pi t)W_2)\ .
 $$
 Then the one and two dimensional cases follow by Theorem~1.1. In
the following we discuss the higher dimensional case.

 Let us fix a wedge $W_0$ and consider the two-dimensional
precosheaf on the translated of $W_0$. Applying Theorem~1.1 we get
a two-parameter group of translations $T(x)$, $x\in\Re^2$, such that
$T(x)\A(W_0)T(x)^*=\A(W_0+x)$ and
$\D_{W_0}^{it}T(x)\D_{W_0}^{-it}=T(\Theta(2\pi t)x)$, where
$\Theta(t)$ is the matrix
 $\left(\matrix{\cosh t&\sinh t\cr\sinh t&\cosh t\cr}\right)$.
 Then, since the algebra of any double cone may be translated into
$\A(W_0)$ by a translation $T(x)$ for some $x$ in $\Re^2$, weak
modular covariance implies strong modular covariance namely, for
all $\O\in\K$,
 $$
\D_{W_0}^{it}\A(\O)\D_{W_0}^{-it}=\A(\L_{W_0}(2\pi t)\O)\ .
 $$
 Now the results in [\rfr(BGLo2)] gives a representation $U$ of the
universal covering of $\Ppo$ such that any $U(g)$ implements a
precosheaf map. Then, according to [\rfr(GuLo2)], $J_W$ has the
correct commutation relations with the modular unitaries and this
implies that $U$ is indeed a representation of $\Ppo$ and extends
to an (anti)-unitary representation of $\Pp$.\endProof

\begremark{1.6} The same argument used in Lemma~2.5 in
[\rfr(BGLo2)] shows that either $U$ is injective or its kernel is
$\Ppo$. Indeed, in order to obtain Corollary~1.5 from weak modular
covariance we had to assume the vacuum vector to be cyclic for the
algebras associated with double cones, therefore the kernel of $U$
cannot consist of the translations only. This case would be allowed
if we assumed strong modular covariance and cyclicity of $\Q$ for
the wedge algebras only, and can be interpreted as a theory on a
different space-time [\rfr(BGLo3)].
 \np
 Now we compare Corollary~1.5 with Theorem~3.2 in [\rfr(GuLo2)] and
notice that, for a local precosheaf on wedges, strong modular
covariance is equivalent to weak modular covariance plus essential
duality. Another equivalent formulation is the following:
 $$
\eqalign{
\s_{\A(W)}^t(\A(\O))&=\A(\L_W(2\pi t)\O)\ ,\qquad\O\subset W\cr
\s_{\A(W)'}^t(\A(\O))&=\A(\L_W(-2\pi t)\O)\ ,\qquad\O\subset W'.\cr}
 $$
 Indeed this formulation still involve only the action of the
modular automorphisms on local subalgebras, and implies essential
duality as in [\rfr(BGLo1)]. Moreover, it can easily be adapted
to conformal theories. Given a local net on the double cones of
the $n$-dimensional Minkowski space, we require that
 $$
\eqalign{
\s_{\A(\O_1) }^t(\A(\O_2))&=\A(\L_{\O_1}( 2\pi t)\O_2)
\ ,\qquad\O_2\subset \O_1\cr
\s_{\A(\O_1)'}^t(\A(\O_2))&=\A(\L_{\O_1}(-2\pi t)\O_2)
\ ,\qquad\O_2\subset \O_1'.\cr}
 $$
where the last equation holds whenever $t$ is in a suitable
neighborhood of the origin, namely $\L_{\O_1}(-2\pi s)\O_2$ is
well defined for any $s$ between $0$ and $t$. In this case, the
same arguments used in this section together with the analysis in
Section~1 of [\rfr(BGLo1)] imply that the net extends to a
conformally covariant precosheaf on the double cones of the
universal covering of the Dirac-Weyl compactification of the
Minkowski space.
 \endremark

\titlea{2. PCT invariance and Spin-Statistics relation}

Modular covariance properties play an important role in the proofs
of PCT theorem and Spin and Statistics relation which are given
in [\rfr(GuLo1),\rfr(GuLo2),\rfr(Kuck2),\rfr(GuLo3),\rfr(BGLo3)].
In particular the geometrical meaning of the modular conjugation
associated with wedge algebras is an essential tool. As proved in
Section~1, such a property follows from the weak modular
covariance. In conformal theories such properties need not to be
assumed, since they follow from conformal covariance [\rfr(BGLo1)].

In this section we review the main steps from modular covariance
to PCT and Spin-Statistics results.

\titleb{Poincar\'e covariant theories} Let $\A$ be a precosheaf on
the wedges of the $n$-di\-men\-sio\-nal Minkowski space $M$ of
von~Neumann algebras in $\B(\H)$, $\H$ a separable Hilbert space.
We assume the same properties as in Section~1, namely essential
duality, the existence of a common cyclic vector $\Q$ for the
algebras associated with double cones and weak modular covariance
as stated in equation (1.20).

Then we consider the quasi-local C$^*$algebra $\Anul$ given by the
inductive C$^*$-limit of the algebras associated with double cones,
and localized representations of $\Anul$, namely representations
which are unitarily equivalent to the identity representation if
restricted to the causal complement of any space-like cone.
Therefore, given any space-like cone $\S$, we may identify the
Hilbert space of $\pi$ with $\H$ in such a way that $\pi|_{\A(\S')}$
$=$ $id|_{\A(\S')}$. The resulting representation is called
localized morphism. Localized morphisms are transportable, namely
may be localized in any space-like cone by a unitary conjugation.

If $\r$ is localized in $\S$ $\subset$ $W$, then $\r$ restricted to
$\Anul\cap\A(W)$ extends to and endomorphism $\r_W$ of $\A(W)$.
Two localized morphisms $\r$ and $\s$  are said {\it locally
conjugate} if, once localized in the same space-like cone $\S$,
for any $W\supset\S$ $\r_W$ and $\s_W$ are conjugate as
endomorphisms of the von~Neumann algebra $\A(W)$ [\rfr(Long3)]. If
$\r$ has finite statistics, we shall call {\it global conjugate}
the conjugate in the sense of Doplicher, Haag and Roberts
[\rfr(DHRo1),\rfr(DHRo2)].

A localized morphism $\r$ is (Poincar\'e) covariant if there exists
a positive energy representation $U_\r$ of the universal covering
of the Poincar\'e group such that $\r\cdot U(g)$ $=$ $\ad
U_\r(g)\cdot\r$.
 We recall that if regularity holds for $\A$ (cf. [\rfr(GuLo1)],
Section~5), any localized morphism $\r$ with finite statistics
is Poincar\'e covariant.

We have already shown in Section~1 that the modular conjugation $J_W$
associated with any wedge $W$ gives a (partial) PT operator, namely
implements the space-time reflection w.r.t. the edge of $W$. The
rest of the PCT theorem is contained in the following:

\begtheorem{2.1} Let $\A$ be a precosheaf on the wedges of the
$n$-dimensional Minkowski space verifying essential duality and
weak modular covariance, and let $\r$ be a localized covariant
morphism. Then, if $j$ denotes the modular anti-automorphism
associated with any given wedge, $\ad j\cdot\r\cdot j$ is a local
conjugate of $\r$ and, if $\r$ has finite statistics, it is a
global conjugate of $\r$.\endtheorem

\begProof The local conjugate property is proven in [\rfr(GuLo1)]
and does not depend on the covariance of $\r$, but only on the
geometrical meaning of the modular conjugation. The proof for the
global conjugate property is contained in [\rfr(BGLo3)], and is a
corollary of the analogous statement in [\rfr(GuLo3)] where
positive energy and covariance play a major role.\endProof

Now we restricts to the high dimensional case ($n\geq3)$.If $\r$ is
an irreducible covariant morphism with finite statistics, the
statistics parameter $\l_\r$ [\rfr(DHRo1)] and the spin
$s_\r:=U_\r(2\pi)$ are two scalar quantities. The index statistics
theorem [\rfr(Long2)] states that $|\l_\r|$ $=$ ${\rm
Ind}(\r)^{-1/2}$, where ${\rm Ind}(\r)$ is the Jones index of the
inclusion $\r_W(\A(W))$ $\subset$ $\A(W)$, $W$ containing the
localization region of $\r$. Setting $\l_\r$ $=$ $|\l_\r|\k_\r$,
the following theorem gives the spin-statistics relation.

\begtheorem{2.2} Let $\A$ be a precosheaf on the wedges of the
$n$-dimensional Minkowski space, $n\geq3$, verifying essential
duality and weak modular covariance, and let $\r$ be an irreducible
covariant morphism with finite statistics. Then $\k_\r$ $=$
$s_\r$.\endtheorem

\begProof If $n\geq4$, the proof is contained in [\rfr(GuLo2)] (cf.
also [\rfr(Kuck2)]). Indeed in this case (or when $\r$ is
localized in a double cane) one may construct the Doplicher Roberts
field algebra [\rfr(DoRo1)], and the theorem follows by the
equality between the statistics operator and $U(2\pi)$ on such
algebra. When $n=3$ and $\r$ is localized in a space-like cone, the
previous technique does not apply and we refer to the proof in
[\rfr(BGLo3)], which is a natural extension of the arguments in
[\rfr(GuLo3)].\endProof

\titleb{Conformal theories on $S^1$} Let $\A$ be a local precosheaf
on the intervals of $S^1$ of von~Neumann algebras on a separable
Hilbert space $\H$, where by interval we mean an open non empty
connected subset of $S^1$ such that the interior $I'$ of its
complement is non empty too. Following [\rfr(GuLo3)] we assume
conformal covariance and the existence of a unique conformally
invariant vector $\Q$ cyclic for the algebra generated by the
$\A(I)$'s. These properties imply that both the modular groups
and the modular conjugations have a geometrical meaning.

Then we consider the universal C$^*$-algebra $\ua$ associated
with the precosheaf $\A$ (cf. [\rfr(Fred1)]), identifying the
local algebras with the corresponding subalgebras of $\ua$. It
turns out that the representations of the precosheaf $\A$ are in
one-to-one correspondence with the localized endomorphisms of
$\ua$, namely the endomorphisms $\r$ of $\A$ such that, for some
interval $I$, $\r|_{\A(I')}$ $=$ $id|_{\A(I')}$.

If $\r$ is a localized covariant endomorphism then finite statistics
is equivalent to finite index. When $\r$ is irreducible, the
index-statistics correspondence holds and the univalence $s_\r:=$
$U_\r(2\pi)$ is a well defined complex number of modulus one.

\begtheorem{2.3} Let $\A$ be a conformal precosheaf on $S^1$, and
let $\r$ be a covariant, finite-statistics, irreducible
endomorphism of the universal algebra $\ua$. Then $j\cdot\r\cdot j$
gives a conjugate endomorphism for $\r$, where $j$ is the modular
anti-automorphism associated with an interval $I$, and the
spin-statistics relation holds, namely $\k_\r$ $=$
$s_\r$.\endtheorem

We refer to [\rfr(GuLo3)] for the proof of this statement
(an independent proof of this theorem based on different ideas,
namely the reconstruction of local fields, has been given recently
by J\"orss [\rfr(Jors2)]).

Further generalizations of the techniques developed in
[\rfr(GuLo3)] are contained in [\rfr(BGLo3)], where spin-statistics
relations for conformal theories on higher-dimensional space-times
or for theories on a different space-time are proven.

 \references
 \end